\documentclass[prl,twocolumn,superscriptaddress,tight,floatfix]{revtex4}%

\usepackage{amsfonts}
\usepackage{amsmath}
\usepackage{amssymb}
\usepackage{graphicx}%

\begin{document}

\title{Field-dependent diamagnetic transition in magnetic superconductor $\rm{Sm_{1.85}Ce_{0.15}CuO_{4-y}}$}

\author{R.~Prozorov}%
\affiliation{Loomis Laboratory of Physics, University of Illinois at Urbana - Champaign, 1110 West
Green Street, Urbana, IL
61801.}%
\affiliation{Department of Physics \& Astronomy and NanoCenter, University of South Carolina, 712
Main St, Columbia, SC 29208.}

\author{D.D. Lawrie}%
\affiliation{Loomis Laboratory of Physics, University of Illinois at Urbana - Champaign, 1110 West
Green Street, Urbana, IL 61801.}

\author{I.~Hetel}%
\affiliation{Regroupement qu\'{e}b\'{e}cois sur les mat\'{e}riaux de pointe, D\'{e}partement de
physique, Universit\'{e} de Sherbrooke, Sherbrooke, Qu\'{e}bec, Canada, J1K 2R1.}

\author{P.~Fournier}%
\affiliation{Regroupement qu\'{e}b\'{e}cois sur les mat\'{e}riaux de pointe, D\'{e}partement de
physique, Universit\'{e} de Sherbrooke, Sherbrooke, Qu\'{e}bec, Canada, J1K 2R1.}

\author{R.~W.~Giannetta}%
\affiliation{Loomis Laboratory of Physics, University of Illinois at Urbana - Champaign, 1110 West
Green Street, Urbana, IL 61801.}



\begin{abstract}
The magnetic penetration depth of single crystal $\rm{Sm_{1.85}Ce_{0.15}CuO_{4-y}}$ was measured
down to 0.4 K in dc fields up to 7 kOe. For insulating $\rm{Sm_2CuO_4}$, Sm$^{3+}$ spins order at
the N\'{e}el temperature, $T_N = 6$ K, independent of the applied field. Superconducting
$\rm{Sm_{1.85}Ce_{0.15}CuO_{4-y}}$ ($T_c \approx 23$ K) shows a sharp increase in diamagnetic
screening below $T^{\ast }(H)$ which varied from 4.0 K ($H = 0$) to 0.5 K ($H =$ 7 kOe) for a field
along the c-axis. If the field was aligned parallel to the conducting planes, $T^{\ast }$ remained
unchanged. The unusual field dependence of $T^{\ast }$ indicates a spin freezing transition that
dramatically increases the superfluid density.
\end{abstract}

\date{4 June 2004}
%

\maketitle

The coexistence of magnetism and superconductivity has been studied in many materials. These
include ternary compounds $\rm{RRh_4B_4}$ and $\rm{RMo_6S_8}$ \cite{maple},  rare-earth (R)
borocarbides, $\rm{RNi_2B_2}C$ \cite{lynn,gammel} and hole-doped copper oxides.\cite{maple3}
$\rm{Nd_{1.85}Ce_{0.15}CuO_{4-y }}$ (NCCO) \cite{maple2,ncco} and
$\rm{Sm_{1.85}Ce_{0.15}CuO_{4-y}}$ (SCCO)\cite{dilichaouch,sumarlin} are widely studied
electron-doped copper oxides in which rare earth magnetic ordering coexists with superconductivity.
Heat capacity measurements have shown peaks at $T_N(\rm{Nd}^{3+}) \approx 1.2$ K \cite{markert} and
$T_N(\rm{Sm}^{3+}) \approx 5$ K \cite{cho,hetel}, respectively. Neutron scattering confirmed that
insulating $\rm{Sm_2CuO_4}$ exhibits Sm$^{3+}$ antiferromagnetism below $T_{N,Sm} \approx 6$ K on
top of the high temperature N\'{e}el ordering of the Cu spins (at $T_{N,Cu} \sim 270$ K). Within
each plane Sm$^{3+}$ spins are ferromagnetically aligned along the c-axis, but with their direction
alternating from one plane to the next.\cite{sumarlin} For this reason, SCCO with its particular
magnetic structure, has been proposed as a possible "$\pi $-phase" superconductor in which the
order parameter changes sign from one CuO layer to the next.\cite{Baladie}

In this Letter, we report measurements of the magnetic penetration depth in
SCCO for magnetic fields applied perpendicular and parallel to the conducting ab-plane. A sharp
increase in diamagnetic screening is observed upon cooling below a temperature $T^{\ast }$ which is
slightly less than the ordering temperature for Sm$^{3+}$ spins. $T^{\ast }$ is rapidly suppressed
by a c-axis magnetic field. The unusual field dependence of $T^{\ast }$ indicates a spin freezing
transition, possibly of Cu$^{2+}$, which in turn enhances the superfluid density.

\begin{figure}[htbp]
\centerline{\includegraphics[width=8.5cm]{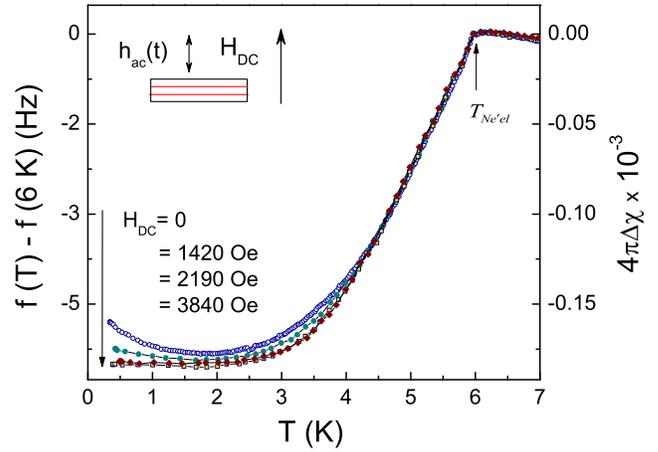}}%
\caption{Frequency shift and susceptibility of insulating $\rm{Sm_2CuO_4}$. AC and DC fields are
parallel to c-axis.$T_N$ is independent of $H_{DC}$ but
the upturn beginning near 4.5 K is suppressed by the field.}%
\label{fig1}%
\end{figure}

The single crystals of SCCO were prepared using a directional flux growth technique.\cite{Peng}
Penetration depth measurements are performed with a 12 MHz tunnel oscillator used previously in
several studies.\cite{Carrington,Prozorov} A dc field up to 7 kOe can be applied along $h_{ac}(t)$
and up to 800 Oe perpendicular to $h_{ac}(t)$. The oscillator frequency shift is proportional to
the sample magnetic susceptibility, $\chi _{m }$, with a sensitivity of $4 \pi \Delta \chi _m
\approx 10^{-7}$ for typical high-$T_c$ crystals ($1 \times 1 \times 0.05$ mm$^3$). In the
superconducting state $-4\pi \chi _m = \left[ 1 - \left(2 \lambda /d\right)
\tanh{(d/2\lambda)}\right]$ where $\lambda$ is the penetration depth and $d$ the sample thickness
\cite{Prozorov}

For the reference, we first measured insulating $\rm{Sm_2CuO_4}$ single crystal which exhibits
Sm$^{3+}$ antiferromagnetism below $T_{N,Sm} \approx 6$ K. Figure~\ref{fig1} shows the frequency
shift and susceptibility for both ac and dc magnetic fields applied along the c-axis. $T_{N,Sm}$ is
insensitive to the applied dc field. For this non-superconducting crystal, the magnitude of the
frequency shift is roughly 5Hz. The susceptibility below 4.5 K is field sensitive, showing an
upturn below 2 K that is suppressed by a c-axis field. The origin of this upturn is not yet
understood, but neutron scattering data in NCCO has shown a similar upturn below
$T_{N}$.\cite{lynn2}

\begin{figure}[htbp]%
\centerline{\includegraphics[width=8.5cm]{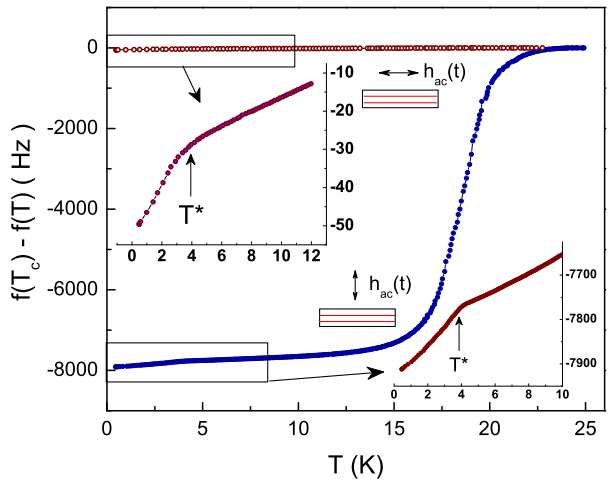}}%
\caption{Frequency shift versus temperature for AC field parallel to ab planes (top) and parallel
to c-axis (bottom). The \textbf{inserts magnify} the regions near $T^{\ast }$ where enhanced diamagnetism occurs.}%
\label{fig2}%
\end{figure}

Doping with Ce$^{4+}$ leads to a semiconductor with a slightly reduced $T_{N,Sm}$.\cite{Jiang}
Subsequent oxygen reduction of $\rm{Sm_{1.85}Ce_{0.15}CuO_4}$ yields the electron-doped
superconductor $\rm{Sm_{1.85}Ce_{0.15}CuO_{4-y}}$ ($T_c \approx$ 23 K). \cite{Peng, Jiang}
Fig.~\ref{fig2} shows the frequency shift in superconducting SCCO for both ac field orientations.
$h_{ac}(t)$ applied along the c-axis generates ab-plane supercurrents. In this case the resonator
senses the ab-plane penetration depth $\lambda _{ab}$ as shown by the bottom curve.  The expanded
region below 6 K shows a drop in frequency below $T^{\ast } \approx 4 $ K corresponding to
\textit{enhanced diamagnetism}. $T^{\ast}$ ranged from $4 - 4.3$ K depending upon the sample. Only
samples with $T_{c} > 20 $ K showed the drop in frequency at $T^{\ast}$.  Two crystals with $T_c
\approx 16$ K showed only a slight break at 4 K. The extra frequency shift of $\approx 100$ Hz is
much larger than the change observed in Fig.~\ref{fig1}.  The top curve in Fig.~\ref{fig2} shows
the frequency shift with $h_{ac}(t)$ along the ab plane. In this orientation the signal is
dominated by very weak interplane supercurrents and the sample is almost magnetically transparent.
Demagnetization corrections are negligible in this orientation and using $-4\pi \chi _m = \left[ 1
- \left(2\lambda_C /d\right) \tanh{(d/2\lambda_C)}\right]$, we estimate the interplane penetration
depth,  $\lambda _{C}$(0) $\approx 400~\mu$m. The inset shows that the diamagnetic transition at
$T^{\ast}$ is also observed for this orientation.

The drop in frequency below $T^{\ast }$ corresponds to $\Delta \lambda _{ab}=\lambda_{ab}$($T^{\ast
}$)-$\lambda_{ab}$ (0.35 K) = 1 $\mu$m, which is very large for a copper oxide material. Reversible
magnetization measurements on aligned powders of SCCO yielded $\lambda _{ab}(0) \approx
0.46~\mu$m.\cite{almassan}. For comparison, $\lambda _{ab}(0) \approx 0.2-0.3~\mu$m in the related
compounds, $\rm{Pr_{1.85}Ce_{0.15}CuO_{4-y}}$ (PCCO) \cite{Prozorov2} and
$\rm{Nd_{1.85}Ce_{0.15}CuO_{4-y }}$ (NCCO).\cite{ncco0}  $\lambda_{ab }$ may be larger in SCCO than
in NCCO and PCCO due to spin fluctuations above the magnetic ordering temperature
\cite{schachinger}. From the top inset of Fig.~\ref{fig2} we estimate that $\Delta \lambda
_{C}=\lambda _{C}(T^{\ast })-\lambda _{C}(0.35~\rm{K}) \approx 60~\mu$m, although the very weak
screening rounds the transition at $T^{\ast }$ considerably. The diamagnetic enhancements shown in
Fig.~\ref{fig2} cannot arise from the additive contribution of the Sm$^{3+}$ spin susceptibility.

Using data from the insulating state, Fig.~\ref{fig1}, we estimate $\chi_{spin}^\parallel \left(
{T_N } \right)-\chi_{spin}^\parallel \left( {0.4\;K} \right) \approx 1.4 \times 10^{-4}$, which
would correspond to a drop in frequency of 3.5 Hz for the sample in Fig.~\ref{fig2}. This estimated
shift is much smaller than the measured 120 Hz shown in the lower inset of Fig.~\ref{fig2}. In
addition, for our superconducting crystals, $\chi_{spin}^\parallel$ is shielded by supercurrents to
within a surface layer of order $\lambda_{ab}$, rendering any additive spin contribution
unobservably small. For $h_{ac}(t)$ along the ab-plane, the field penetrates most of the sample and
would excite the bulk spin susceptibility $\chi _{spin}^\bot$. However, $\chi _{spin}^\bot $ is
nearly temperature independent while the upper inset of Fig.~\ref{fig2} shows a 20 Hz drop in
frequency.  Misalignment of the sample axes could mix contributions from $\lambda _{C}$ and
$\lambda _{ab}$.  We estimate the maximum error from misalignment to be 0.2 Hz, which is far
smaller than the drop shown in Fig.~\ref{fig2}.   The drop in frequency is also far too large to be
explained by any plausible amount of magnetostriction.

The penetration depth measured in a resonator is enhanced by the susceptibility of magnetic ions:
$\lambda _{meas}=\lambda (1 + 4\pi \chi _{spin})^{1/2}$ where $\lambda$ is the penetration depth
that would exist in their absence.\cite{ginzburg,Cooper} This effect explains the upturn in
$\lambda _{ab }$ observed in $\rm{Nd_{1.85}Ce_{0.15}CuO_{4-y }}$ where the Nd$^{3+}$ moments are
large and remain paramagnetic to much lower temperatures.\cite{prozorov3}  For SCCO, this effect
would give $\lambda _{ab}(T_N) - \lambda_{ab}(0.4~\textrm{K}) \approx 10^{-3} \lambda_{ab}(T_N)$
which is far too small to account for the drop in frequency observed.  A somewhat similar situation
was observed in $\rm{ErNi_2B_2C}$.\cite{gammel} Those authors also concluded that the drop in
$\lambda$ at T$_{\rm{N\acute{e}el}} = 6$ K could not be attributed to the $\sqrt{\mu_{\rm{spin}}}$
factor.

\begin{figure}[htbp]%
\centerline{\includegraphics[width=7.5cm]{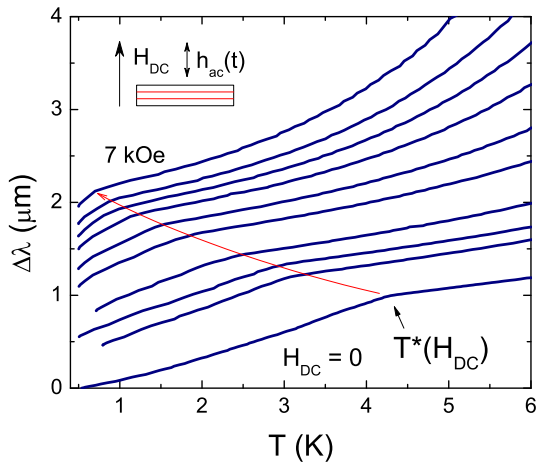}}%
\caption{Change in penetration depth with both AC and DC fields along c-axis, for values of
$H_{DC}$ ranging from 0 to 7 kOe.  The line through the data plots indicates $T^{\ast}$ as a
function of
$H_{DC}$.  The curves are offset from influence of Campbell vortex penetration depth.}%
\label{fig3}
\end{figure}

Penetration depth \cite{Hou} and Josephson critical current measurements \cite{Vaglio} in
$\rm{SmRh_4B_4}$ have shown enhanced superfluid density below $T_{N}$, consistent with theories of
s-wave, antiferromagnetic superconductors.\cite{Rama,Chi} The situation is likely to be quite
different in SCCO.

\begin{figure}[htbp]%
\centerline{\includegraphics[width=8cm]{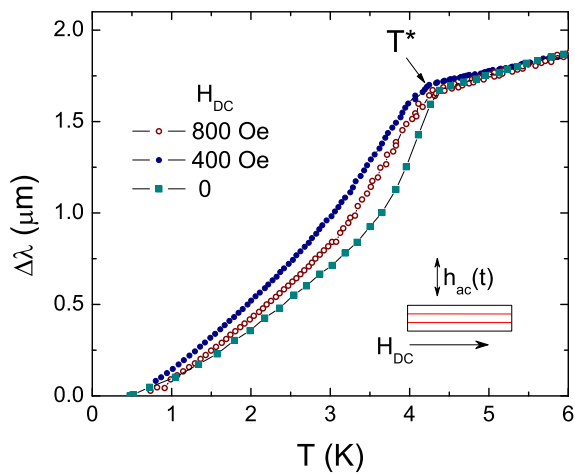}}%
\caption{$\Delta \lambda (T)$ for AC field still along c axis but H$_{dc}$ along ab planes.
$T^{\ast }$ is unchanged by the magnetic field.}%
\label{fig4}
\end{figure}

Fig.~\ref{fig3} shows the effect of a static magnetic field $H_{DC}$ on $\lambda_{ab }$. Both
$h_{ac}(t)$ and $H_{DC}$ were applied along the c-axis. In marked contrast to the insulating phase,
$T^{\ast }$ in the superconductor drops rapidly with field, reaching 0.6 K for H = 0.7 Tesla.
Fig.~\ref{fig4} shows the effect of orienting $H_{DC}$ parallel to the conducting planes, but
maintaining $h_{ac}(t)$ along the c-axis. The large drop in $\lambda_{ab}$ remains, and $T^{\ast}$
is unchanged.

\begin{figure}[htbp]%
\centerline{\includegraphics[width=7.5cm]{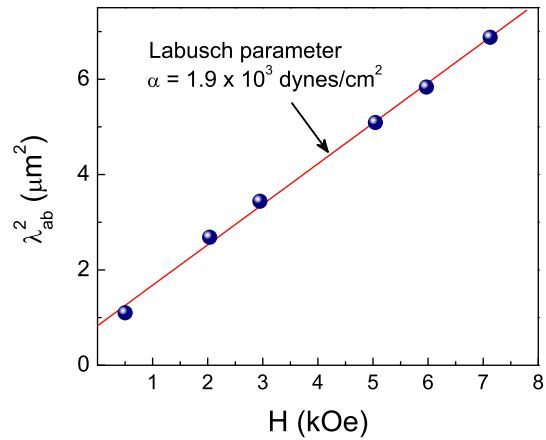}}%
\caption{Square of the penetration depth versus applied field at $T = 0.5$ K}%
\label{fig5}
\end{figure}

The field dependence of $\lambda_{ab}$ is plotted in Fig.~\ref{fig5}, where data at T = 0.5 K have
been taken directly from Fig.~\ref{fig3}.  The plot shows classic vortex behavior in which
$\lambda^2 \left( H \right) = \lambda_{London}^2  + \phi_{0}H/(4 \pi \alpha)$.  The second term on
the right is the square of the Campbell\cite{Campbell} pinning depth where $\alpha$ is the
Labusch\cite{Labusch} pinning constant. $\lambda^2(H,T = 0.5~ \textrm{K})$ is linear in H and thus
dominated by vortex motion with a pinning constant of $\alpha$ = $1.9 \times 10^3$ dynes/$cm^2$, a
value roughly three orders of magnitude smaller than observed in typical YBCO crystals.\cite{Wu}.
This weak pinning is consistent with recent magneto-optical measurements performed on similar SCCO
samples \cite{promo} and may result from the spin polarization in the vortex core. In
$\rm{DySo_6S_8}$ ($T_c = 1.6$ K, $T_N = 0.4$ K) for example, Dy spins apparently assume an
antiferromagnetic alignment outside the vortex core and a spin-flop orientation
inside.\cite{Krzyszton}  Most important, Fig.~\ref{fig5} demonstrates that the response below
$T^{\ast}$ still derives from superconductivity and not from spins located in possibly
non-superconducting regions.

\begin{figure}[htbp]%
\centerline{\includegraphics[width=7.5cm]{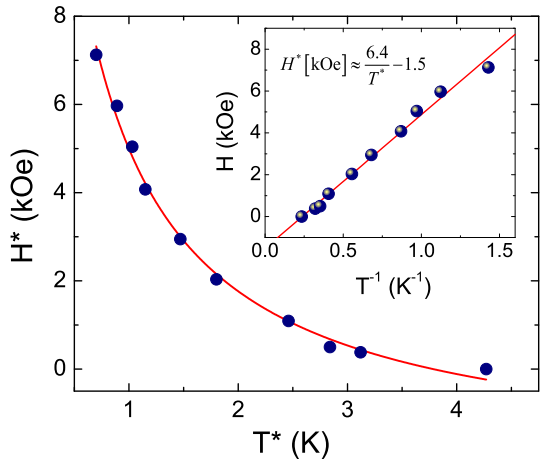}}%
\caption{Location of the diamagnetic transition $T^{\ast }$(H) in the H-T
plane. Inset shows the data plotted versus inverse temperature.}%
\label{fig6}
\end{figure}

The most striking feature of the data is the strong field dependence of  $T^{\ast}$, taken from
Fig.~\ref{fig3} and plotted in Fig.~\ref{fig6}. For an antiferromagnet with $T_{N}(H=0) = 6$ K,
\textbf{a} 0.7 Tesla field might reduce $T_{N}$ by 0.1 K at most.\cite{Foner} Random field effects,
possibly from Ce-induced disorder, can increase the field dependence. \cite{Aharony} However, heat
capacity measurements in superconducting Sm$_{1.85}$Ce$_{0.15}$CuO$_{4 }$ showed that the peak
appearing at $T_{N,Sm} \sim 5$ K \cite{cho,hetel} is insensitive to fields as large as
9T\cite{hetel}, ruling out this scenario. A superconducting impurity phase would exhibit a strong
field dependence, but any such phase would require a transition temperature of 4 K and a strong
critical field anisotropy to explain the difference between Figs.~\ref{fig3} and \ref{fig4}.
Finally, there is good evidence that the Sm-Sm exchange constant is unaffected by superconductivity
in the layers.  A careful study of $T_{N}$ with various dopants showed that Ce doping was most
effective in lowering $T_{N}$ but subsequent oxygen reduction, required for superconductivity, had
a negligible effect.\cite{Jiang} It appears that another ingredient beyond the Sm spins is required
to explain Fig.~\ref{fig6}.

The inset to Fig.~\ref{fig6} shows a fit to the expression  $H\left( \rm{Tesla} \right) =
0.64/T^{\ast} - 0.15$.  Precisely this functional dependence was reported for the spin freezing
transition line of $\alpha - Fe_{92}Zr_{8}$.\cite{Ryan}. In this frustrated Heisenberg ferromagnet,
transverse spin components undergo a field dependent spin-glass transition at $T_{xy}$, far below
the temperature for longitudinal spin ordering. We conjecture that the boundary line in
Fig.~\ref{fig6} represents a change in superfluid density caused by a similar spin freezing
transition.  Of the e-doped superconductors measured ($\rm{Pr_{1-x}LaCe_{x}CuO_{4-y}}$,
$\rm{Nd_{1.85}Ce_{0.15}CuO_{4+y}}$ and SCCO), only SCCO shows a transition to enhanced diamagnetism
at low temperatures.  This observation suggests that ordering of the Sm$^{3+}$ spins changes the
magnetic environment and initiates a freezing transition of the Cu$^{2+}$ spins.

Spin freezing in conventional superconductors has been studied theoretically and leads to a
strongly reduced thermal smearing in the density of states and in some cases the opening of a gap.
\cite{nass,schachinger} Evidence for Cu$^{2+}$ spin freezing in e-doped cuprates has come from
$\mu$SR measurements in over-oxygenated $\rm{Nd_{1.85}Ce_{0.15}CuO_{4+y}}$, where Cu spins are
found to undergo a spin-glass transition at 4-5 K \cite{Lascialfari}. There is also evidence for a
spin-glass region in the $\rm{Pr_{1-x}LaCe_{x}CuO_{4-y}}$.\cite{Kuroshima} No such studies have
been reported for SCCO. While spin freezing has been extensively studied in
$\rm{La_{1-x}Sr_{x}CuO_{4}}$ \cite{Kumagai} this is the first evidence, to our knowledge, of its
influence on the superfluid density in either a conventional or unconventional superconductor.

Recent neutron scattering experiments on NCCO also suggest that Cu$^{2+}$ spins are involved.  A
c-axis magnetic field showed no significant effect on the magnetic ordering of Cu$^{2+}$ in either
insulating $\rm{Nd_2CuO_4}$ or semiconducting $\rm{Nd_{1.85}Ce_{0.15}CuO_{4}}$.  However, the field
had a large effect on Cu$^{2+}$ ordering in superconducting
$\rm{Nd_{1.85}Ce_{0.15}CuO_{4-y}}$.\cite{Kang,Matsuura}  Evidently Cu magnetic order in an
electron-doped cuprate is far more responsive to a magnetic field once the system is doped
sufficiently to superconduct: a result entirely consistent with our data.

In conclusion, superconducting $\rm{Sm_{1.85}Ce_{0.15}CuO_{4-y}}$ shows a strong enhancement of
diamagnetic screening below $T^{\ast }$ = 4 K. $T^{\ast }$ is rapidly suppressed with a c-axis
field, suggesting a freezing transition for Cu$^{2+}$ spins.

We wish to thank A.~Aharony, L.~N.~Bulaevskii, A.~I.~Buzdin, E.~Chia, R.~L.~Greene,
P.~J.~Hirschfeld, J.~W.~Lynn, D.~Morr, C.~Panagopolous, M.~Poirier, M.~B.~Salamon, C.~M.~Varma, and
M.~B.~Weissman for many useful discussions. Work at UIUC was supported through NSF DMR 01-01872.
Work at USC was supported by the NSF/EPSCoR under Grant No. EPS-0296165. PF acknowledges the
support of CIAR, CFI, NSERC, FQRNT and the Universit\'{e} de Sherbrooke.


\begin{thebibliography}{99}
\bibitem{maple} \textit{Superconductivity in Ternary Compounds},
ed. M.B. Maple and O. Fischer (Springer-Verlag, Berlin, 1982)

\bibitem{lynn}J.~W. Lynn {\it et al.}, Phys. Rev. B \textbf{55}, 6584 (1997)

\bibitem{gammel}P.~L. Gammel {\it et al.}, Phys. Rev. Lett. \textbf{82}, 1756 (1999)

\bibitem{maple3}J.~T.~Markert {\it et al.}, \textit{Physical Properties of High Temperature Superconductors-I}, pp.
265-337, ed. D.M. Ginsberg, (World Scientific, 1989)

\bibitem{maple2}M.~B. Maple {\it et al.} Physica C \textbf{162-164}, 296 (1989)

\bibitem{ncco} P.~Fournier, E.~Maiser and R.~L.~Greene, NATO ASI Series, Series B: Physics \textbf{371}, 145-158 (1998)

\bibitem{dilichaouch}Y.~Dalichaouch {\it et al.}, Phys. Rev. Lett. \textbf{64}, 599 (1990)

\bibitem{sumarlin}I.~W.~Sumarlin {\it et al.}, Phys. Rev. Lett. \textbf{68}, 2228 (1992)

\bibitem{markert}J.~T.~Markert {\it et al.}, Physica C \textbf{158}, 178 (1989)

\bibitem{cho}B.~K.~Cho {\it et al.}, Phys. Rev. B \textbf{63}, 214504 (2001)

\bibitem{hetel}I.~Hetel, M.~Poirier, and P.~Fournier, unpublished (2004)

\bibitem{Baladie}I.~Baladie {\it et al.}, Phys. Rev. B \textbf{63}, 054518 (2001);
A.V. Andreev, A.I. Buzdin and R.M. Osgood, JETP Lett. \textbf{52}, 55 (1990)

\bibitem{Peng}J.~L.~Peng, Z. Y. Li, and R. L. Greene, Physica C \textbf{177}, 79 (1991)

\bibitem{Carrington}A.~Carrington {\it et al.}, Phys. Rev. B \textbf{59}, R14173 (1999)

\bibitem{Prozorov}~R.~Prozorov {\it et al.}, Phys. Rev. B \textbf{62}, 115 (2000)

\bibitem{lynn2}J.~W.~Lynn{\it et al.}, Phys. Rev. B, \textbf{41}, 2569 (1990).

\bibitem{Jiang}B.~Jiang {\it et al.} , Phys. Rev. B \textbf{45}, 2311 (1992)

\bibitem{almassan}C.~C.~Almasan {\it et al.}, Phys. Rev. B \textbf{45}, 1056 (1992)

\bibitem{Prozorov2}R.~Prozorov {\it et al.}, Appl. Phys. Lett. \textbf{77}, 4202 (2000)

\bibitem{ncco0}S.~M.~Anlage {\it et al.}, Phys. Rev. \textbf{50}, 523 (1994); A. A. Nugroho {\it et al.}, Phys. Rev. B \textbf{60}, 15384 (1999)

\bibitem{schachinger}E.~Schachinger, W. Stephan, and J. P. Carbotte, Phys. Rev. B \textbf{37}, 5003 (1988)

\bibitem{ginzburg}V.~L.~Ginzburg, Zh. Eksp. Teor. Fiz. \textbf{31}, 202 (1956), Sov. Phys. JETP \textbf{4}, 153 (1957)

\bibitem{Cooper}J.~R.~Cooper, Phys. Rev. B \textbf{54}, R3753 (1996)

\bibitem{prozorov3}R.~Prozorov {\it et al.}, Phys. Rev. Lett. \textbf{85}, 3700 (2000)

\bibitem{Hou}M.~K.~Hou {\it et al.} , Sol. St. Comm. \textbf{65}, 895 (1988)

\bibitem{Vaglio}R.~Vaglio{\it et al.}, Phys. Rev. Lett. \textbf{53}, 1489 (1984)

\bibitem{Rama}T.~V.~Ramakrishnan and C.~M. Varma , Phys. Rev. B \textbf{24}, 137 (1981)

\bibitem{Chi}H.~Chi and A.D.S. Nagi , Jour. Low Temp. Phys. \textbf{86}, 139 (1992)

\bibitem{Campbell}A.~M.~Campbell, J. Phys. C \textbf{2}, 1492 (1969); \textbf{4}, 3186 (1971)

\bibitem{Labusch}R.~Labusch, Phys. Rev. \textbf{170}, 470 (1968)

\bibitem{Wu}D-H.~Wu and S. Sridhar, Phys. Rev. Lett. \textbf{65}, 2074 (1990)

\bibitem{promo}R.~Prozorov, A. Snezhko and P. Fournier, Physica C \textbf{405}, 265 (2004)

\bibitem{Krzyszton}T.~Krzyszton and K.~Rogacki, Eur. Phys. J. B \textbf{30}, 181 (2002)

\bibitem{Foner} Y.~Shapira {\it et al.}, Phys. Rev. Lett. \textbf{23}, 98 (1969)

\bibitem{Aharony}S.~Fisher and A. Aharony, J. Phys. C \textbf{12}, L729 (1979)

\bibitem{Ryan} D.~H.~Ryan {\it et al.}, Phys. Rev. B \textbf{63}, 140405 (2001)

\bibitem{nass} M.J. Nass {\it et al.}, Phys. Rev. B \textbf{23}, 1111 (1981)

\bibitem{Lascialfari}A.~Lascialfari, P. Ghigna, and F. Tedoldi, Phys. Rev. B \textbf{68}, 104524 (2003)

\bibitem{Kuroshima}S.~Kuroshima {\it et al.}, Physica C \textbf{392}, 216 (2003)

\bibitem{Kumagai}K.~Kumagai {\it et al.}, Physica C \textbf{235-240}, 1715 (1994)

\bibitem{Kang} H.~J.~Kang {\it et al.}, Nature (London) \textbf{423}, 522 (2003)

\bibitem{Matsuura}M.~Matsuura {\it et al.}, Phys. Rev. B \textbf{68}, 144503 (2003)

\end{thebibliography}
\end{document}